\newcommand{\be}{\begin{equation}}
\newcommand{\ee}{\end{equation}}
\newcommand{\ba}{\begin{array}}
\newcommand{\ea}{\end{array}}
\newcommand{\bea}{\begin{eqnarray}}
\newcommand{\eea}{\end{eqnarray}}

\newcommand{\vbar}{\raisebox{.17ex}{\rule{.04em}{1.35ex}}}
\newcommand{\vbarind}{\raisebox{.01ex}{\rule{.04em}{1.1ex}}}
\newcommand{\R}{\ifmmode {\rm I}\hspace{-.2em}{\rm R} \else ${\rm I}\hspace{-.2em}{\rm R}$ \fi}
\newcommand{\T}{\ifmmode {\rm I}\hspace{-.2em}{\rm T} \else ${\rm I}\hspace{-.2em}{\rm T}$ \fi}
\newcommand{\N}{\ifmmode {\rm I}\hspace{-.2em}{\rm N} \else \mbox{${\rm I}\hspace{-.2em}{\rm N}$} \fi}
\newcommand{\B}{\ifmmode {\rm I}\hspace{-.2em}{\rm B} \else \mbox{${\rm I}\hspace{-.2em}{\rm B}$} \fi}
\newcommand{\Hil}{\ifmmode {\rm I}\hspace{-.2em}{\rm H} \else \mbox{${\rm I}\hspace{-.2em}{\rm H}$} \fi}
\newcommand{\C}{\ifmmode \hspace{.2em}\vbar\hspace{-.31em}{\rm C} \else \mbox{$\hspace{.2em}\vbar\hspace{-.31em}{\rm C}$} \fi}
\newcommand{\Cind}{\ifmmode \hspace{.2em}\vbarind\hspace{-.25em}{\rm C} \else \mbox{$\hspace{.2em}\vbarind\hspace{-.25em}{\rm C}$} \fi}
\newcommand{\Q}{\ifmmode \hspace{.2em}\vbar\hspace{-.31em}{\rm Q} \else \mbox{$\hspace{.2em}\vbar\hspace{-.31em}{\rm Q}$} \fi}
\newcommand{\Z}{\ifmmode {\rm Z}\hspace{-.28em}{\rm Z} \else ${\rm Z}\hspace{-.28em}{\rm Z}$ \fi}

\documentclass[10pt,conference,a4paper]{IEEEtran}
\usepackage{cite}
\usepackage{graphicx}
\usepackage[cmex10]{amsmath}
\usepackage{url}
\usepackage{algorithm}
\usepackage{amssymb}
\usepackage{tabularx}
\usepackage{float}
\usepackage{fancybox}
\usepackage{longtable}
\usepackage{verbatim}
\usepackage{flushend}
\IEEEoverridecommandlockouts
\graphicspath{{./Kuvat/}}

\begin{document}


\title{System Level Performance Evaluation of LTE-V2X Network}
\author{\begin{minipage}[t]{1.0\linewidth}\centering Petri Luoto$^\star$, Mehdi Bennis$^\star$, Pekka Pirinen$^\star$, Sumudu Samarakoon$^\star$, Kari Horneman$^\dagger$, Matti Latva-aho$^\star$\end{minipage}\hspace{-1mm} \\
\begin{minipage}[t]{0.34\linewidth}\centering $^\star$Centre for Wireless Communications\\University of Oulu, Finland\\P.O. Box 4500, FI-90014 Oulu \{petri.luoto, bennis, pekka.pirinen, sumudu, matti.latva-aho\}@ee.oulu.fi\end{minipage}
\begin{minipage}[t]{0.33\linewidth}\centering $^\dag$Nokia\\Kaapelitie 4, \\ P.O. Box 319, FI-90620 Oulu\\ kari.horneman@nokia.com\end{minipage}
\thanks{This research was supported by the Finnish Funding Agency for Technology and Innovation (TEKES), Nokia, Anite, Huawei
Technologies, and Infotech Oulu Graduate School. Kari Horneman from Nokia earns special thanks for invaluable
guidance for this study.}}

\maketitle

\sloppy

\begin{abstract}
Vehicles are among the fastest growing type of connected devices. Therefore, there is a need for Vehicle-to-Everything (V2X) communication i.e. passing of information from a Vehicle-to-Vehicle (V2V) or Vehicle-to-Infrastructure (V2I) and vice versa. In this paper, the main focus is on the communication between vehicles and road side units (RSUs) commonly referred to as V2I communication in a multi-lane freeway scenario. Moreover, we analyze network related bottlenecks such as the maximum number of vehicles that can be supported when coverage is provided by the Long Term Evolution Advanced (LTE-A) network. The performance evaluation is assessed through extensive system-level simulations. Results show that new resource allocation and interference mitigation techniques are needed in order to achieve the required high reliability requirements, especially when network load is high.

\end{abstract}


\section{Introduction}\label{sec:intro}
Vehicle-to-Everything (V2X) communication has a crucial role for enabling reliable and low latency services for vehicles such as forward collision warning, road safety services and emergency stop \cite{3GPP_V2X_3}. Vehicles are among the fastest growing type of connected devices \cite{D2D_1}. Therefore, there is a need for better solutions for V2X communication, especially if Long Term Evolution Advanced (LTE-A) is used. V2X communication is an ongoing 3GPP Study item and LTE-V2X is an important feature of LTE Release 14 \cite{3GPP5}.

Because most of the V2X applications are real-time, strict requirements are needed \cite{METISD1.1}. The end-to-end latency requirements of less than 5 ms for message sizes of about 1600 bytes need to be guaranteed for all V2X transmissions with a probability of 99.999\%. Traffic is either event-driven or periodically sent, with a typical time interval of 100 ms. Relative speeds up to 500 km/h should be supported in highway scenarios.

The importance of V2X communication has been around for years because it is considered as an important part of future Intelligent Transportation Systems (ITS). Dedicated Short-Range Communications (DSRC) has been around for a decade and it is based on the IEEE 802.11p technology, which appeared as the most promising technique for V2X communication \cite{V2X_IEEE}, \cite{V2X_IEEE2}. However, recent studies have preferred using LTE as the V2X technology \cite{V2X_LTE}, \cite{V2X_LTE2}, mainly because LTE cellular network infrastructure already exists \cite{V2X_LTE3}. Aforementioned studies have been focusing mostly on Vehicle-to-Vehicle (V2V) communication or analyzing transmission delays.

In this paper, we focus on the Vehicle to Infrastructure (V2I) based communication in the downlink direction, i.e., infrastructure is used to send messages to vehicles for example informing that new route should be selected because an accident has happened on the highway. Here, the performance of a six-lane highway, which is covered by LTE-A RSU (road side unit) network is studied. We consider the LTE-A RSU network without Proximity Service (ProSe) capability. The performance is evaluated by using an LTE-A compliant system level simulator, where RSUs are serving vehicles in the highway in the downlink direction. The aim is to analyze network related problems such as the maximum number of vehicles that can be supported by the network and outage probabilities in the network.

This paper is organized as follows. The system and link models are defined in Sections \ref{sec:model_system} and \ref{sec:model_link}, respectively. Section \ref{sec:results} provides the performance evaluation of the vehicular network. Finally, Section \ref{sec:conclusion} concludes the paper.

\section{System Model} \label{sec:model_system}
A network with single user single-input multiple-output (SU-SIMO) and with single user multiple-input multiple-output (SU-MIMO) transmission schemes are compared with orthogonal frequency-division multiple access (OFDMA). Let $\mathcal{B}$ be a set of RSU where RSU $b$ has $N_\text{t}$ transmit antennas (Tx), which serves a set of vehicles $\mathcal{V}_b$, where vehicle $v$ has $N_\text{r}$ receive antennas (Rx). The frequency domain consists of a set $\mathcal{S}$ of subcarriers.

In the SU-SIMO transmission scheme the signal vector received from the RSU $b$ by the vehicle $v \in \mathcal{V}_b$ over the subcarrier $s \in \mathcal{S}$ can be written as
\begin{equation}
\textbf{y}_{b,v}^s = \textbf{h}_{b,v}^s\textbf{x}_{b,v}^s+\sum_{i\in\mathcal{B}\setminus b}\textbf{{h}}_{i,v}^s\textbf{x}_{i,v}^s+\textbf{n}_{b,v}^s\label{receivedsignal},
\end{equation}
where $\textbf{x}_{b,v}^s\in\mathbb{C}^{N_{t}}$ is the transmitted signal from the desired RSU $b$ to vehicle $v$ over subcarrier $s$, $\textbf{h}_{b,v}^s\in\mathbb{C}^{N_{r}\times N_{t}}$ is the channel vector from desired RSU $b$ to the $v$th vehicle over the $s$th subcarrier, $\textbf{x}_{i,v}^s \in\mathbb{C}^{N_{t}}$ is the transmitted signal from the $i$th interfering RSU at subcarrier $s$, $\textbf{h}_{i,v}^s$ is the channel vector from the $i$th interfering RSU to the $v$th vehicle at subcarrier $s$, and $\textbf{n}_{b,v}^s\sim\mathcal{C}\mathcal{N}(0,N_0\textbf{I}_{N_{r_{v}}})$ denotes the additive noise with zero mean.

We analyze the performance of the network considering two types of receivers: maximum ratio combining (MRC) and linear minimum mean square error (LMMSE). The MRC weight vector $\textbf{w}_{b,v}^s\in\mathbb{C}^{N_{r}\times{N_{t}}}$ is given by
\begin{equation}
\textbf{w}_{b,v}^s = (\textbf{h}_{b,v}^s)^*,
\end{equation}
where $(\cdot)^*$ denotes the conjugate transpose.

When LMMSE and $N_t = 2$ used, LTE specific precoder providing the best performance has been applied in transmission. In the SU-MIMO transmission scheme, the received signal vector from RSU $b$ by the vehicle $v$ over the subcarrier $s$ is given by
\begin{equation}
\textbf{y}_{b,v}^s = \textbf{H}_{b,v}^s\textbf{x}_{b,v}^s+\sum_{i\in\mathcal{B}\setminus b}\textbf{{H}}_{i,v}^s\textbf{x}_{i,v}^s+\textbf{n}_{b,v}^s\label{receivedsignal2},
\end{equation}
where $\textbf{H}_{b,v}^s\in\mathbb{C}^{N_{r}\times N_{t}}$ is the channel matrix from desired RSU $b$ to the \emph{v}th vehicle over the \emph{s}th subcarrier, and $\textbf{{H}}_{i,v}^s$ is the channel matrix from the \emph{i}th interfering RSU to the \emph{v}th vehicle at subcarrier \emph{s}.

For the LMMSE filter, the weight matrix $\textbf{W}_{b,v}^s\in\mathbb{C}^{N_{r}\times{N_{t}}}$ of the LMMSE receiver is given by
\begin{equation}
\textbf{W}_{b,v}^s = \text{arg}~\underset{\textbf{W}_{b,v}^s}{\text{min}}\text{E}[\|\textbf{x}_{b,v}^s-\hat{\textbf{x}}_{b,v}^s\|^2],
\end{equation}
where $\hat{\textbf{x}}_{b,v}^s = (\textbf{W}_{b,v}^s)^H\textbf{y}_{b,v}^s$ is the vector of estimated received data. Therefore, the weight matrix can be written as \cite{TSE2005}
\begin{equation}
\textbf{W}_{b,v}^s = (\textbf{H}_{b,v}^s(\textbf{H}_{b,v}^s)^H+\textbf{R}_{b,v}^s)^{-1}\textbf{H}_{b,v}^s,
\end{equation}
where $\textbf{R}_{b,v}^s$ is the inter-cell interference plus noise covariance matrix and it is assumed to be known at the receiver.


\section{Link Model} \label{sec:model_link}
\begin{figure}
  \centering
    \includegraphics[trim = 0mm 0mm 0mm -4mm, clip, width=0.5\textwidth]{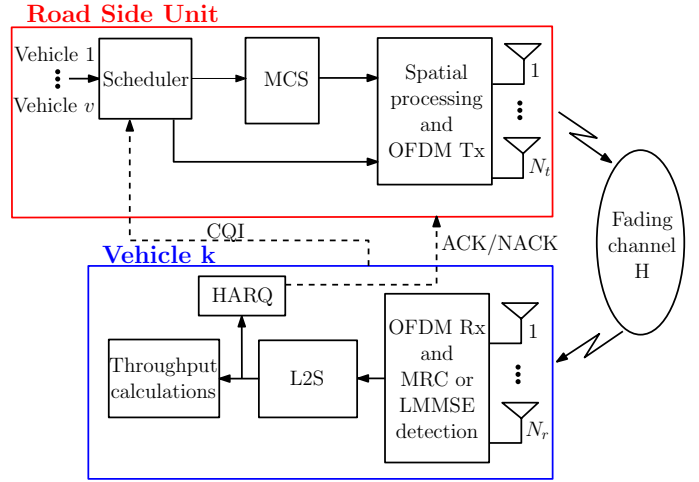}
    \caption{\label{link_model}Block diagram of the link model.}
\end{figure}

The link model between an RSU and a vehicle is illustrated in Fig. \ref{link_model}. A detailed link-to-system interface (L2S) is used in the simulations. Each vehicle is then paired to an RSU based on path loss model
\begin{equation}
\text{PL}_\text{dB} = 100.7 +23.5\text{log}_{10}(d),
\end{equation}
where $d$ is distance in kilometers. The path loss model is given in \cite{3GPP_pathloss}, where it is described as macro to relay path-loss model.

A geometry-based stochastic channel model (GSCM) \cite{WINII,3GPP3} is used to model fast fading and shadowing losses for all links. Channel parameters are determined stochastically, based on the statistical distributions extracted from channel measurements \cite{itur2135}. Stochastic channel parameters are adopted from urban macro environment.

The link model starts from the scheduler that is responsible for resource allocation for vehicles. Throughout the simulations proportional fair (PF) scheduling is used. The scheduler utilizes channel-quality indicator (CQI) information transmitted by the vehicle. Based on the CQI information resource allocation is performed. The CQI provides information to the RSU about the link adaptation parameters. In the simulator, CQI is estimated from the received signal and for each vehicle signal-to-interference-plus-noise ratio (SINR) is calculated for every physical resource block (PRB). In order to model a practical closed loop system, periodic and delayed CQI is assumed.

After scheduling, modulation and coding scheme (MCS) selection is performed for scheduled vehicles. Finally, before the data is sent over the fading channel, transmitter side spatial and OFDM processing are performed. The cyclic prefix is assumed to be longer than the multipath delay spread, and thus inter-symbol-interference is not considered.

At the receiver, perfect frequency and time synchronization is assumed. Link-to-system mapping is performed using mutual information effective SINR mapping (MIESM) \cite{MIESM}. This significantly reduces the computational overhead compared with exact modeling of the radio links, while still providing sufficiently accurate results. In the link-to-system interface, SINR is calculated and it is mapped to corresponding average mutual information. Based on the MIESM value, the frame error probability (FEP) is approximated according to a predefined frame error rate (FER) curve of the used MCS. Based on the FER, successful and erroneous frames can be detected, and hybrid automatic repeat request (HARQ) can take the control of retransmissions. An acknowledgement (ACK) or a negative acknowledgement (NACK) message is sent back to the RSU to signal the success or failure of the transmission, respectively. The results are obtained by simulating a predefined number of channel samples.


\section{System Level Performance Results}\label{sec:results}
System level simulations are particularly useful for studying network related issues, such as resource allocation, interference management and mobility management. In this work, LTE-A system level simulator is used to model a highway RSU network which is used to serve vehicles at high speed.

\begin{figure}
  \centering
    \includegraphics[width=0.5\textwidth]{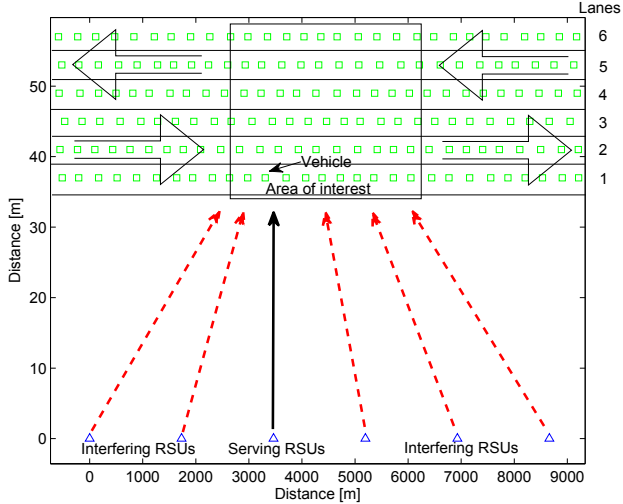}
    \caption{\label{Road_network}Highway layout where RSUs are located next to the road.}
\end{figure}

The simulator uses a six lane high way layout as shown in Fig. \ref{Road_network}. Vehicles on lanes 1 to 3 are moving to the right and vehicles on lanes 4 to 6 are moving to the left. The RSU network is along the road, 35 m from the highway. Although all RSUs and vehicles are modeled, the performance analysis is conducted for the middle RSUs. This models the wrap-a-round type of interference in the system.

Vehicles are deployed to lanes and distance between vehicles is in the range from 38 m to 300 m. This parameter is used to model different vehicle densities in the network \cite{3GPP_V2X_1}. Each RSU unit serves multiple vehicles and one vehicle is connected to a single RSU. Generally, V2X traffic is periodically sent. However, for this work we have mapped 1600 byte package which is sent with a time interval of 100 ms to be equivalent to a constant transmission with a target rate of 128 kb/s per vehicle. Table \ref{params} summarizes the main simulation parameters and assumptions which are used through simulations.

\begin{table}
\centering
\caption{Simulator parameters and assumptions.}%
\label{params}
\begin{tabular}{|p{3.3cm}|p{4cm}|}
  \hline
  \textbf{Parameter} & \textbf{Assumption} \\ \hline
      Duplex mode & FDD \\ \hline
      System bandwidth & 10 MHz \\ \hline
      Number of PRBs & 50 \\ \hline
Antenna configuration & 1 Tx $\times$ 2 Rx,  \\
                   & 2 Tx $\times$ 2 Rx \\ \hline
      Vehicle speed & 140 km/h \\ \hline
      Inter vehicle distances & min 38 meters and max 300 meters \\ \hline
      Inter RSU distance & 1732 m \\ \hline
      Receivers & MRC \\ \hline
      HARQ & Chase combining \\ \hline
      Transmission power & 46 dBm \\ \hline
      Feedback CQI period & 6 ms \\ \hline
      Feedback CQI delay & 2 ms \\ \hline
      Channel estimation & Ideal \\ \hline
      Network synchronization & Synchronized \\ \hline
      Receiver type & MRC or LMMSE \\ \hline
      L2S interface metric & MIESM \\ \hline
      Traffic model & Continuous constant rate transmission\\ \hline
      Scheduler & Proportional fair \\ \hline
      Target rate & 128 kb/s\\
  \hline
\end{tabular}
\end{table}

\subsection{Performance Analysis}
\begin{figure}
  \centering
    \includegraphics[width=0.45\textwidth]{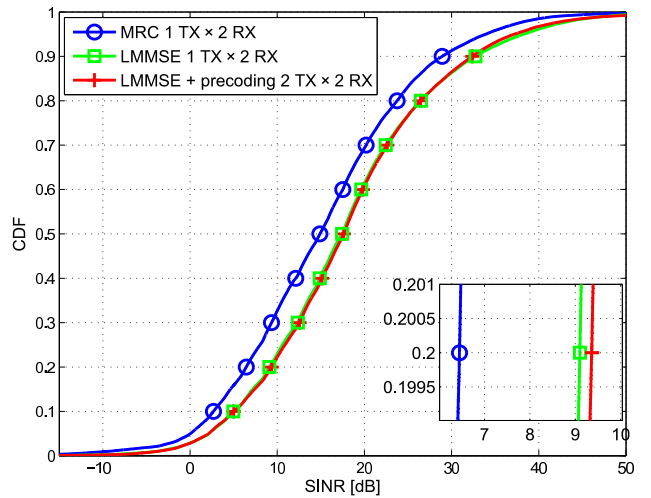}
    \caption{\label{CDF_SINR}SINR comparison of the different receiver algorithms with varying inter vehicle distances from 200 meters up to 300 meters.}
\end{figure}

In Fig. \ref{CDF_SINR} cumulative distribution function (CDF) of the SINR of the V2I network is analyzed where the minimum and maximum distances between vehicles are 200 m and 300 m, respectively which corresponds to around 40 vehicles are connected to each RSU. The SINR characteristics for the dense scenario is similar to Fig. \ref{CDF_SINR} because the number of interference sources (RSUs) remains unchanged. Fig. \ref{CDF_SINR} shows that about 50\% of vehicles can achieve an SINR of 15 dB and cell edge vehicles (5\% from CDFs) can achieve SINR of 2 dB. The difference in SINR between MRC and MMSE is 3 dB and the gain from precoding is minimal, around 0.3 dB. Based on this result, the main challenge is to serve the vehicles at the cell edge, requiring high number of PRBs due to lower SINR than vehicles at cell center with high SINR. Furthermore, the problem is even greater when the network becomes dense due to the increased number of cell edge vehicles.

\begin{figure}
  \centering
    \includegraphics[width=0.5\textwidth]{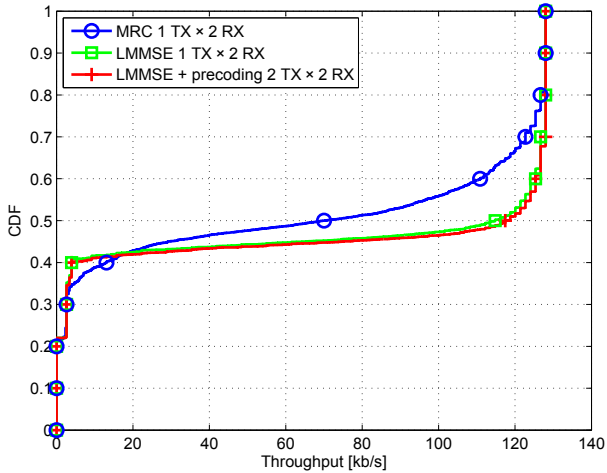}
    \caption{\label{CDF_THR1}Throughput comparison of the different receiver algorithms with varying inter vehicle distances from 38 meters up to 116 meters.}
\end{figure}

\begin{figure}
  \centering
    \includegraphics[width=0.5\textwidth]{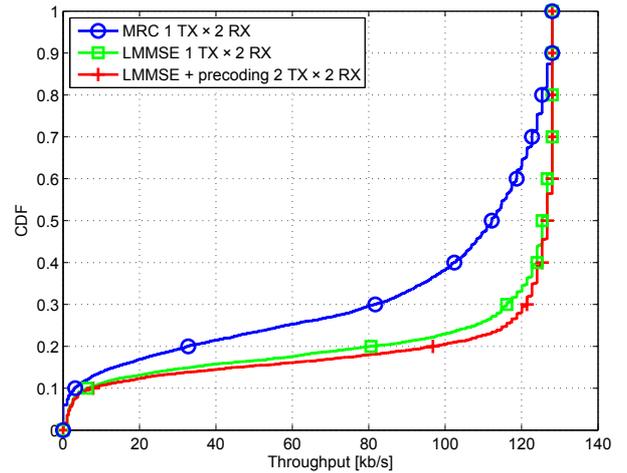}
    \caption{\label{CDF_THR2}Throughput comparison of the different receiver algorithms with inter vehicle distance of 116 meters.}
\end{figure}

\begin{figure}
  \centering
    \includegraphics[width=0.5\textwidth]{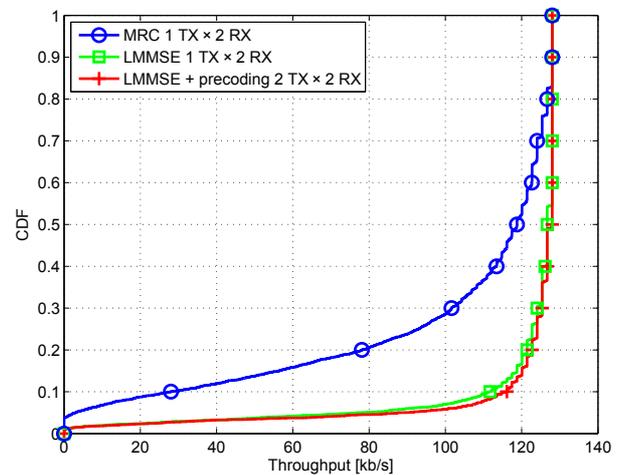}
    \caption{\label{CDF_THR3}Throughput comparison of the different receiver algorithms with varying inter vehicle distances from 100 meters up to 200 meters.}
\end{figure}

\begin{figure}
  \centering
    \includegraphics[width=0.5\textwidth]{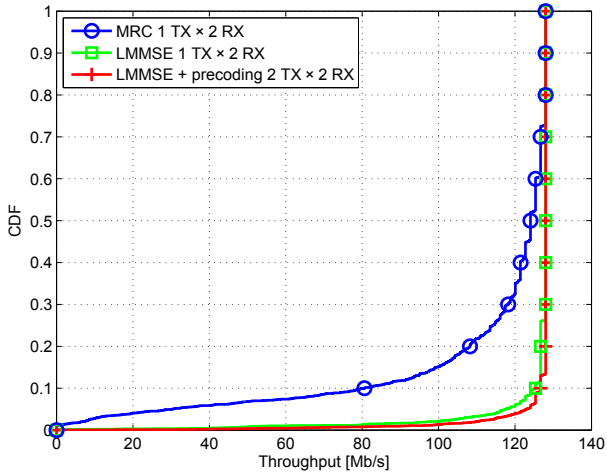}
    \caption{\label{CDF_THR4}Throughput comparison of the different receiver algorithms with varying inter vehicle distances from 200 meters up to 300 meters.}
\end{figure}

Next, the throughput of the V2I network is analyzed for different vehicle densities as shown in Figs. \ref{CDF_THR1} -- \ref{CDF_THR4}. In Fig. \ref{CDF_THR1}, having the distance between vehicles from 38 m to 116 m, almost 40\% of the vehicles are in outage and only 20\% can achieve the target rate. In this scenario, around 135 vehicles are connected to each RSU. This is the most extreme scenario where vehicles are very close to each other. When 10 MHz bandwidth is used it corresponds to 50 PRBs which is insufficient to serve all 135 vehicles. The RSU needs additional 85 PRBs to serve all 135 vehicles by allocating one PRB per vehicle. Another option is to maintain a mean rate of 128 kb/s by scheduling. For example, if data rate from one resource block is 384 kb/s, a vehicle needs a new PRB every third time slot in order to achieve the mean rate. We have modified the PF scheduler to support this aforementioned method. The main problem is the cell edge vehicles because their data buffer is cumulatively increasing. Furthermore, when the PF scheduler is serving cell edge vehicles, a high number of other vehicles are not served, which leads to overall low performance of the network.

In Fig. \ref{CDF_THR2}, when each vehicle is at safe distance, 116 m apart from each other (around 90 vehicles are connected to each RSU) performance is improved drastically and outage probability is around 10\%. In this scenario, vehicles can benefit from LMMSE receiver, and precoding further improves the performance. However, outage probability is too high, because there are too many vehicles in the network and they cannot be served.

Fig. \ref{CDF_THR3} shows the throughput when density is further decreased (around 65 vehicles are connected to each RSU). In this scenario, with the MRC receiver only 5\% of the vehicles are in outage. However, this is not acceptable, especially if data transmission priority is 0, which is the highest level \cite{3GPP_V2X_2}. With the LMMSE receiver, the cell edge (5\% from CDFs) throughput is around 80 kb/s. Even though the number of vehicles is low the PF scheduler cannot guarantee the target rate of 128 kb/s for every vehicle in the network.

Fig. \ref{CDF_THR4} shows the performance when the network is sparse, around 40 vehicles are connected to each RSU. When LMMSE receiver with the precoding is used, 86.5\% of the vehicles can achieve the target rate of 128 kb/s. Table \ref{probability} illustrates probabilities to achieve the target throughput with the different receiver algorithms and inter vehicle distances. For a dense network, there is only 50\% probability to achieve the target rate. As the network becomes sparse, the probability to achieve the target rate increases and reaches almost 99\%. However, these results indicate the need of novel resource allocation and interference mitigation techniques in order to achieve the probability of 99.999\%, which was required for end-to-end delay \cite{METISD1.1}.

Table \ref{cell_edge} summarizes the throughput of cell edge vehicles for different inter vehicle distances and receiver algorithms. This table shows that cell edge throughput needs to be improved drastically, because in the dense scenario, all the cell edge vehicles are in outage. Based on all the results, it is evident that new resource allocation and interference mitigation techniques are needed in order to achieve the required high reliability percentages especially when the network load is high.

\begin{table}
\centering
\caption{Probability to achieve a target throughput for different inter vehicle distances and receiver algorithms.}%
\label{probability}
\begin{tabular}{|c|c|c|c|}
  \hline
  \textbf{Vehicle distance [m]} & \textbf{MRC} & \textbf{LMMSE} & \textbf{LMMSE + precoding} \\ \hline
     [38 116] & 49.7 & 54.4 & 55.2 \\ \hline
     [116 116] &69.6 & 80.3 & 82.3 \\ \hline
     [100 200] & 78.4 & 93.4 & 94.2 \\ \hline
     [200 300] & 87.9 & 98.1 & 98.9 \\ \hline
\end{tabular}
\end{table}

\begin{table}
\centering
\caption{Cell edge vehicles throughput [kb/s] for different vehicle distances and receivers.}%
\label{cell_edge}
\begin{tabular}{|c|c|c|c|}
  \hline
  \textbf{Vehicle distance [m]} & \textbf{MRC} & \textbf{LMMSE} & \textbf{LMMSE + precoding} \\ \hline
     [38 116] & -- & -- & -- \\ \hline
     [116 116] &-- & 2 & 2 \\ \hline
     [100 200] & -- & 75 & 91 \\ \hline
     [200 300] & 30 & 117 & 123 \\ \hline
\end{tabular}
\end{table}


\section{Conclusion}\label{sec:conclusion}
We have evaluated the performance of Vehicle to Infrastructure (V2I) based communication in a freeway scenario. The
framework has been established under the LTE-A compliant system simulation platform where the system throughput performance and signal-to-interference-plus-noise ratio have been rigorously assessed. In the evaluation, we analyzed whether the existing LTE technologies are sufficient enough to achieve the target performances. Provided numerical results show the challenges related to V2I communication when the network is dense and reliability requirement is high. The results indicate the need of novel resource allocation and interference mitigation techniques to meet the performance requirements. Developing these techniques will be tackled in our future work. Moreover, we will solve Vehicle-to-Everything (V2X)  related communication problems, by considering LTE-A networks with Proximity Service (ProSe) capabilities.


\bibliographystyle{IEEEtran}
\bibliography{IEEEabrv,Referencelist2}

\begin{thebibliography}{10}
\providecommand{\url}[1]{#1}
\csname url@samestyle\endcsname
\providecommand{\newblock}{\relax}
\providecommand{\bibinfo}[2]{#2}
\providecommand{\BIBentrySTDinterwordspacing}{\spaceskip=0pt\relax}
\providecommand{\BIBentryALTinterwordstretchfactor}{4}
\providecommand{\BIBentryALTinterwordspacing}{\spaceskip=\fontdimen2\font plus
\BIBentryALTinterwordstretchfactor\fontdimen3\font minus
  \fontdimen4\font\relax}
\providecommand{\BIBforeignlanguage}[2]{{%
\expandafter\ifx\csname l@#1\endcsname\relax
\typeout{** WARNING: IEEEtran.bst: No hyphenation pattern has been}%
\typeout{** loaded for the language `#1'. Using the pattern for}%
\typeout{** the default language instead.}%
\else
\language=\csname l@#1\endcsname
\fi
#2}}
\providecommand{\BIBdecl}{\relax}
\BIBdecl

\bibitem{3GPP_V2X_3}
\BIBentryALTinterwordspacing
``{3GPP TR 36.885 V0.4.0} {Study on LTE-based V2X Services}.'' [Online].
  Available: \url{http://www.3gpp.org/dynareport/36885.htm}
\BIBentrySTDinterwordspacing

\bibitem{D2D_1}
A.~Khelil and D.~Soldani, ``On the suitability of {D}evice-to-{D}evice
  communications for road traffic safety,'' in \emph{2014 IEEE World Forum on
  Internet of Things (WF-IoT)}, March 2014, pp. 224--229.

\bibitem{3GPP5}
\BIBentryALTinterwordspacing
``{3GPP} {N}ews, {LTE} support for the connected car.'' [Online]. Available:
  \url{http://http://www.3gpp.org/news-events/3gpp-news/1675-lte_automotive}
\BIBentrySTDinterwordspacing

\bibitem{METISD1.1}
\BIBentryALTinterwordspacing
``{METIS} deliverable {D}1.1. {S}cenarios, requirements and {KPIs} for {5G}
  mobile and wireless system.'' [Online]. Available:
  \url{https://www.metis2020.com/documents/deliverables/}
\BIBentrySTDinterwordspacing

\bibitem{V2X_IEEE}
L.~Le, A.~Festag, R.~Baldessari, and W.~Zhang, ``Vehicular wireless short-range
  communication for improving intersection safety,'' \emph{{IEEE} Commun.
  Mag.}, vol.~47, no.~11, pp. 104--110, 2009.

\bibitem{V2X_IEEE2}
R.~Protzmann, B.~Schunemann, and I.~Radusch, ``{The influences of communication
  models on the simulated effectiveness of V2X applications},'' \emph{{IEEE}
  Commun. Mag.}, vol.~49, no.~11, pp. 149--155, 2011.

\bibitem{V2X_LTE}
M.~Botsov, M.~Klugel, W.~Kellerer, and P.~Fertl, ``Location-based resource
  allocation for mobile {D2D} communications in multicell deployments,'' in
  \emph{2015 IEEE International Conference on Communication Workshop (ICCW)},
  2015, pp. 2444--2450.

\bibitem{V2X_LTE2}
C.~Lottermann, M.~Botsov, P.~Fertl, and R.~Mullner, ``Performance evaluation of
  automotive off-board applications in {LTE} deployments,'' in \emph{Vehicular
  Networking Conference (VNC), 2012 IEEE}, Nov. 2012, pp. 211--218.

\bibitem{V2X_LTE3}
\BIBentryALTinterwordspacing
J.~Yoshida, ``{Prelude to 5G: Qualcomm, Huawei Muscle into V2X},'' Oct. 2015.
  [Online]. Available:
  \url{http://www.eetimes.com/document.asp?doc_id=1328030&page_number=2}
\BIBentrySTDinterwordspacing

\bibitem{TSE2005}
D.~Tse and P.~Viswanath, \emph{Fundamentals of Wireless Communication}.\hskip
  1em plus 0.5em minus 0.4em\relax Cambridge, UK: Cambridge University Press,
  2005.

\bibitem{3GPP_pathloss}
\BIBentryALTinterwordspacing
``{3GPP TR 36.814 V1.7.0}. {Further Advancements for E-UTRA Physical Layer
  Aspects (Release 9)}.'' [Online]. Available:
  \url{http://www.3gpp.org/dynareport/36814.htm}
\BIBentrySTDinterwordspacing

\bibitem{WINII}
\BIBentryALTinterwordspacing
``{WINNER II channel models, D1.1.2 V1.0}.'' [Online]. Available:
  \url{http://www.cept.org/files/1050/documents/winner2%20-%20final%20report.pdf}
\BIBentrySTDinterwordspacing

\bibitem{3GPP3}
``{3rd Generation Partnership Project, Technical Specification Group Radio
  Access Network, Spatial channel model for Multiple Input Multiple Output
  (MIMO) simulations, 3GPP Technical report 25.996 v6.1.0}.''

\bibitem{itur2135}
``{Guidelines for evaluation of radio interface technologies for IMT-Advanced,
  Report ITU-R M.2135-1}.''

\bibitem{MIESM}
X.~He, K.~Niu, Z.~He, and J.~Lin, ``Link layer abstraction in {MIMO-OFDM}
  system,'' in \emph{International Workshop on Cross Layer Design (IWCLD)},
  Sep. 2007, pp. 41 --44.

\bibitem{3GPP_V2X_1}
\BIBentryALTinterwordspacing
``{3GPP TSG RAN WG1 Meeting \#82 R1-153800}. {O}verview of {V2X} evaluation
  methodology.'' [Online]. Available:
  \url{http://www.3gpp.org/ftp/tsg_ran/WG1_RL1/TSGR1_82/Docs/}
\BIBentrySTDinterwordspacing

\bibitem{3GPP_V2X_2}
\BIBentryALTinterwordspacing
``{3GPP TSG RAN WG1 Meeting \#83 R1-156609}. {C}onsiderations on {V2X} traffic
  priority and relative resource allocation.'' [Online]. Available:
  \url{https://portal.3gpp.org/ngppapp/CreateTdoc.aspx?mode=view&contributionId=666667}
\BIBentrySTDinterwordspacing

\end{thebibliography}

\end{document}